\documentclass[
    reprint,superscriptaddress,
    bibnotes,
    amsfonts,amsmath,amssymb,floatfix,longbibliography,aps,pre
]{revtex4-2}

\usepackage[dvipdfmx]{graphicx}
\usepackage{amsthm,mathrsfs,dsfont}
\usepackage{mathtools}
\usepackage{braket}
\usepackage{bm}
\usepackage{booktabs}
\usepackage{multirow}
\usepackage[colorlinks,linkcolor=blue,citecolor=blue]{hyperref}

\newcommand{\tr}{\operatorname{Tr}}
\newcommand{\ham}{\hat{H}}
\newcommand{\s}{\hat{\sigma}}

\graphicspath{{./fig/}}  
\newcommand\calA{\mathcal{A}}
\newcommand\calB{\mathcal{B}}

\newcommand\calL{\mathcal{L}}

\newcommand\calP{\mathcal{P}}

\newcommand{\hO}{\hat{O}}

\newcommand{\eref}[1]{Eq.~\eqref{#1}}

\newcommand{\fref}[1]{Fig.~\ref{#1}}

\newcommand{\sref}[1]{Sec.~\ref{#1}}
\newcommand{\aref}[1]{Appendix~\ref{#1}}
\newcommand{\ccite}[1]{Ref.~\cite{#1}}

\allowdisplaybreaks[1]
\usepackage{datetime2}
\begin{document}
\title{Work extractability from energy eigenstates under optimized local operations}

\author{Shotaro~Z.~Baba}
\email{baba.shotaro.s@gmail.com}
\affiliation{Department of Applied Physics, The University of Tokyo, Hongo, Bunkyo-ku, Tokyo 113-8656, Japan}

\author{Nobuyuki Yoshioka}
\email{nyoshioka@ap.t.u-tokyo.ac.jp}
\affiliation{Department of Applied Physics, The University of Tokyo, Hongo, Bunkyo-ku, Tokyo 113-8656, Japan}
\affiliation{Theoretical Quantum Physics Laboratory, RIKEN Cluster for Pioneering Research (CPR), Wako-shi, Saitama 351-0198, Japan}
\affiliation{JST, PRESTO, 4-1-8 Honcho, Kawaguchi, Saitama, 332-0012, Japan}

\author{Takahiro Sagawa}
\affiliation{Department of Applied Physics, The University of Tokyo, Hongo, Bunkyo-ku, Tokyo 113-8656, Japan}
\affiliation{Quantum-Phase Electronics Center (QPEC), The University of Tokyo, Tokyo 113-8656, Japan}
\begin{abstract}
    We examine the relationship between the second law of thermodynamics and the energy eigenstates of quantum many-body systems that undergo cyclic unitary evolution. 
    Using a numerically optimized control protocol,
    we analyze how the work extractability is affected by the integrability of the system.
    Our findings reveal that, in nonintegrable systems the number of work-extractable energy eigenstates converges to zero, even when the local control operations are optimized. 
    In contrast, in integrable systems, there are exponentially many eigenstates from which positive work can be extracted, regardless of the locality of the control operations.
    We numerically demonstrate that such a strikingly different behavior can be attributed to the number of athermal energy eigenstates.
    Our results provide insights into the foundations of the second law of thermodynamics in isolated quantum many-body systems, which are expected to contribute to the development of  quantum many-body heat engines.
\end{abstract}
\maketitle
\section{Introduction}
    \begin{figure}[tbp]
        \centering
        \includegraphics[width=\linewidth]{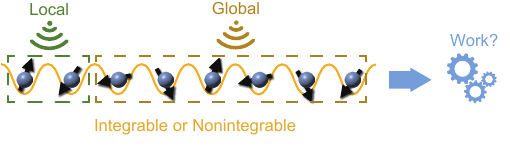}
        \caption{
            Graphical description of the object of this study:
            We consider the work extraction from energy eigenstates of the quantum many-body systems.
            We investigate the dependence of the extraction on the locality of the control operators and integrability of the system.
        }\label{fig:keymap}
    \end{figure}

    The second law of thermodynamics in isolated quantum many-body systems has drawn significant attention in recent years, driven by a crucial question in statistical physics: how does macroscopic irreversibility originate from microscopically reversible dynamics~\cite{Rigol2008-qf,Iyoda2017-lm,Tasaki2000-de,Goldstein2013-cn,Ikeda2015-sj,Kaneko2017-wd}?
    In the context of work extraction, in particular, the exploration of the relationship between the second law of thermodynamics and thermal pure quantum states, which are indistinguishable from the Gibbs ensemble at a macroscopic level, constitutes a fundamental problem in statistical mechanics~\cite{Kaneko2019-ip,Dorner2013-hp,Gallego2014-yn,Perarnau-Llobet2016-zv,Modak2017-fz,Le2018-jr}.

    One of the most well-known expressions of the second law of thermodynamics is embodied in Planck's principle, which claims that it is impossible to extract work from the Gibbs state via adiabatic cycles.
    Correspondingly, it is understood to be unattainable to extract work from canonical ensemble through any cyclic operations, which is called the passivity of the Gibbs ensemble~\cite{Pusz1978-oq,Lenard1978-nf,mitsuhashi2022characterizing}.
    In contrast, there does not exist any no-go principle to prevents work extraction from pure quantum states; the second law can be violated if one is able to perform arbitrary unitary operations with arbitrary precision. 
    Note that this does not contradict the eigenstate thermalization hypothesis~\cite{Neumann1929-qs,Deutsch1991-bu,Srednicki1994-du,Rigol2008-qf, kim_testing_2014,beugeling_finite-size_2014,yoshizawa_numerical_2018,steinigeweg_eigenstate_2013,sorg_relaxation_2014,khodja_relevance_2015,mondaini_eigenstate_2016}, since it only states the macroscopic indistinguishability from the thermal equilibrium.

    Despite the great amount of effort by the existing works, our understanding on the relationship between the second law of thermodynamics and pure quantum many-body states is limited, in particular when all control operations are subject to local constraint (See Fig.~\ref{fig:keymap}).    
    A prior study by Ref.~\cite{Kaneko2019-ip} has investigated the connection between the work extractability from single energy eigenstates that undergo a simple quench dynamics. 
    Ref.~\cite{Kaneko2019-ip} reveals that, in a nonintegrable system, it is impossible with quench dynamics to extract work from any energy eigenstate corresponding to a positive temperature,
    while we can extract work from an exponentially large number of eigenstates if both the initial and the quenched Hamiltonians are integrable.
    While this previous work shows that the integrability of the model yields a qualitative difference in the work extractability, the control operation is limited to a simple class of quench dynamics. 
    In order to address more general situations, one needs to carefully examine the entire degrees of freedom in the control operations.

    \begin{table}[tb]
        \centering\begin{tabular}{lcc}
            \toprule
                \multirow{2}{*}{Control locality}&\multicolumn{2}{l}{Initial Hamiltonian \(\hat{H}(0)\)}\\
                \cmidrule(lr){2-3}
                &Nonintegrable&Integrable\\
            \midrule
                Local& -&
                \(O(\exp(cL))\) \\
                Non-local& \(O(\exp(cL))\)&
                \(O(\exp(cL))\)\\
            \bottomrule
        \end{tabular}
        \caption{
                The size scaling of $D_{\rm pos},$ the count of work-extractable energy eigenstates within a fixed energy shell that corresponds to a positive temperature.
                The count $D_{\rm pos}$ converges to zero under large system size in nonintegrable systems if the control operation is local, while it grows exponentially in other cases. 
            }\label{teb:outliers_scaling}
    \end{table}
    
    In the present work, we analyze the work extractability of energy eigenstates  subject to numerically optimized cyclic control operations. 
    As we summarized in Table~\ref{teb:outliers_scaling}, for nonintegrable systems, the number of work-extractable states corresponding to positive temperatures converges to zero even with the optimized protocol if the control operations are local, 
    while we find that there are exponentially many work-extractable states when control operations are global.
    In sharp contrast,  integrable systems allow work extraction from exponentially many energy eigenstates, whether the control operations are local or global.
    We further numerically demonstrate that such a difference in work extractability can be attributed to the distribution of the entanglement entropy (EE); integrable systems have exponentially many athermal states so that we may reduce the energy of the state without increasing the EE, which is a prohibited scenario in nonintegrable systems since almost all eigenstates are expected to be thermal.
    
    The remainder of the paper is organized as follows. 
    In Sec.~\ref{sec:setup}, we describe the setup and the algorithms to optimize the protocols extracting work. 
    In Sec.~\ref{sse:result},  we present our main numerical results.
    Finally, we give the conclusion and discussion in Sec.~\ref{sse:conclusion}.

\section{Setup} \label{sec:setup}
    \subsection{Work extraction from energy eigenstates}\label{sss:work_extraction}
        We present the definition of work extractable eigenstates assuming cyclic control operations. 
        Let $\hat{H}$ be a Hamiltonian that satisfies $\hat{H}  |E_\alpha\rangle = E_\alpha |E_\alpha\rangle$ where the $\alpha$-th eigenstate with energy $E_\alpha$ is denoted as $|E_\alpha\rangle$.
        Under some unitary time evolution $U(t)$, we define the work extraction from the $\alpha$-th eigenstate at time $t$ as
        \begin{eqnarray}
            W_\alpha(t):&=& \tr\left[\hat{H} \rho_\alpha(0) \right] - \tr \left[ \hat{H}\rho_\alpha(t)\right] \\
            &=& E_\alpha - \tr\left[ \hat{H} U(t) |E_\alpha \rangle \langle E_\alpha | U^\dag(t)\right],
        \end{eqnarray}
        where we have denoted the time evolved eigenstate as $\rho_\alpha(t) := U(t)\ket{E_\alpha} \bra{E_\alpha}U^\dag(t)$.
        In particular, we assume that the unitary $U$ is discretized as
        \begin{align}
            U(t)=\prod_{n=1}^{N_t}\exp (-i\delta t\ham(n\delta t)),
        \end{align}
        where $\hat{H}(t)$ is a time-dependent Hamiltonian that controls the dynamics of the system.
        Here, the time step is homogeneously given as $\delta t$, and $N_t=t/\delta t$ is the number of discrete steps.
        In practice, we consider an implementable set of local operations $\mathcal{B} = \{\hat{O}_i\}$ and take their linear combination to constitute the time-dependent Hamiltonian as
        \begin{eqnarray}
            \hat{H}(t) = \sum_{\hat{O}_i\in\mathcal{B}} \gamma_i(t) \hat{O}_i,\label{eq:greedy_ham}
        \end{eqnarray}
        where $\gamma_i(t)$ denotes the coefficient of the $i$-th operator at time $t$.

        We also introduce a metric to measure the work extractability. We indicate the number of eigenstates in some energy shell from which positive work is extracted as
        \begin{align}
            D_{\mathrm{pos}}(t):=\left|\{\ket{E_{\alpha}}~|~W_{\alpha}(t)\geq \varepsilon L,~E_\alpha \in {\rm shell}\}\right|,\label{eq:dout}
        \end{align}
        where the threshold $\varepsilon$ is introduced for the purpose of numerical stability.

    \subsection{Optimization of control operations}\label{sss:optimization}
        In order to quantify the performance of control protocols, we introduce a reward function. As an example of a smooth function that explicitly rewards $W_\alpha > \varepsilon L$, we define the following:
        \begin{align}
            r(t):=\sum_{E_\alpha\in\mathrm{shell}}\sigma_{a}(w_{\alpha}(t)-\varepsilon)+c(w_{\alpha}(t)-\delta)\theta(\delta-w_{\alpha}(t)),\label{eq:reward}
        \end{align}
        where \(w_{\alpha}(t):=W_{\alpha}(t)/L\) is the work density with \(L\) being the system size, \(\sigma_{a}(x):=1/(1+\exp (-ax))\) is a sigmoid function, \(\theta(x)\) is the unit step function, and \(c,\varepsilon,\delta\) are some hyperparameters that determine the behavior of the reward function.
        Concretely, \(a\) in the sigmoid function controls the width of  its ascending segment,
        while \(c\) regulates the slope of the linear function so that work below a certain threshold \(\delta\) is penalized. 
        In the following, we fix the hyperparameters as \(a=30, c=0.1,\varepsilon=0.15,\delta=0.3\). 
        While we expect that the main result is not  affected significantly by the choice of the reward function, we leave this as a future work.
        
        To reveal the work extractability under optimal control techniques, we consider two optimization methodologies: 
        the gradient-based algorithm
        and the deep reinforcement learning (RL) algorithm. 
        In the gradient-based algorithm we compute the gradient $\partial r/\partial \gamma_i$ with a constraint such that the Frobenius norm of the control Hamiltonian is bounded as $\|\hat{H}(t)\|_2 \leq C$, where the norm upper bound is fixed as $C=\sqrt{2 L d}$ with $d=2^L$ being the full Hilbert space dimension. 
        As an implementable operation set, we consider $\mathcal{B}_k$ as a set of translationally invariant and spatial inversion-symmetric operators that act on at most $k$ contiguous sites.
        On the other hand, the RL algorithm constructs a strategy to choose a unitary from a discrete set of unitaries \(\left\{ U_{m} \right\}\) at each time step, so that we can maximize the discounted reward expectation value that is estimated using a deep neural network. 
        In this work, \(\left\{ U_{m} \right\}\) is generated from  \(\calB_{2}\). 
        See \aref{app:alg} for details of the optimization methods.

\section{Main Results}\label{sse:result}
    \begin{figure}[tbp]
        \centering
        \includegraphics[width=\linewidth]{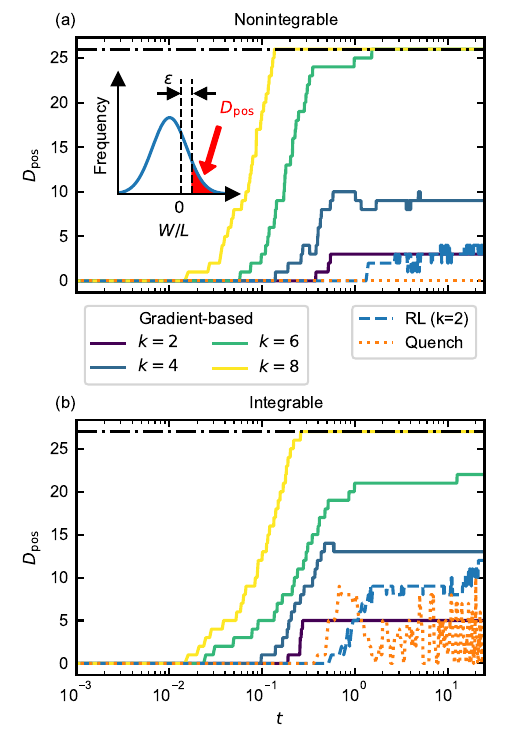}
        \caption{
            The time evolution of $D_{\rm pos}$ generated by control protocols optimized by gradient-based method, RL, and simple quench.
            The initial and final Hamiltonians are (a)~nonintegrable and (b)~integrable.
            The results by the gradient-based method with various operation set $\mathcal{B}_k$ are shown by real lines, while the results from the RL-optimized protocol and simple quench dynamics are shown by blue dashed and orange dotted lines, respectively.
            Note that the RL algorithm performs optimization of the protocol corresponding to \(k=2\).
            The black dash-dotted lines represent the number of eigenstates in the given energy shell.
            The system size is \(L=12\).
            The threshold is taken as \(\varepsilon=0.15\), which is identical to the width of the energy shell (see \aref{app:threshold_dep} for the dependence on the choice of \(\varepsilon\)). 
            The quench dynamics is performed under the Hamiltonian~\eqref{eq:ising} with $(h,g)=(0, 1.5)$, which is also employed in Ref.~\cite{Kaneko2019-ip}.
        }\label{fig:Dout_t}
    \end{figure}
    \begin{figure}[tbp]
        \centering
        \includegraphics[width=\linewidth]{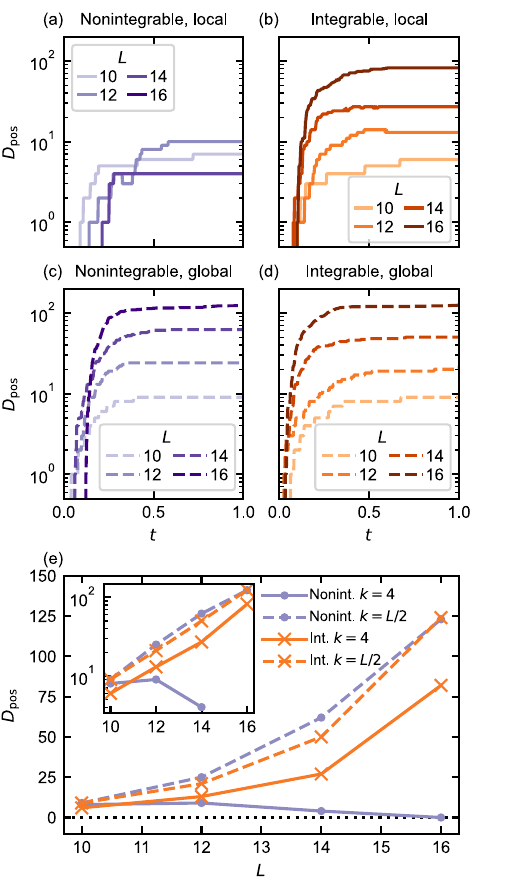}
        \caption{
            The system-size scaling of $D_{\rm pos}$ generated by the control protocol optimized by the gradient-based method.
            (a)--(d)~The dynamics of \(D_{\mathrm{pos}}\) during the protocol for \(L=10,12,14,16\).
            The integrability of the initial/final Hamiltonians and the locality of the operation are 
            (a)~nonintegrable and local~(\(k=4\)), (b)~integrable and local~(\(k=4\)), (c)~nonintegrable and global~(\(k=L/2\)), and (d)~integrable and global~(\(k=L/2\)).
            The absence of the line for \(L=16\) in panel~(a) represents that there is no energy eigenstate from which positive work is extracted.
            The scaling of $D_{\rm pos}(t)$ at $t=1$ is summarized in (e).
            The inset of panel~(e) shows that the $D_{\rm pos}(t)$ increase exponentially with the system size except for the nonintegrable system under local control.
        }\label{fig:Dout_locality}
    \end{figure}

    \begin{figure*}[tbp]
        \centering
        \includegraphics[width=\linewidth]{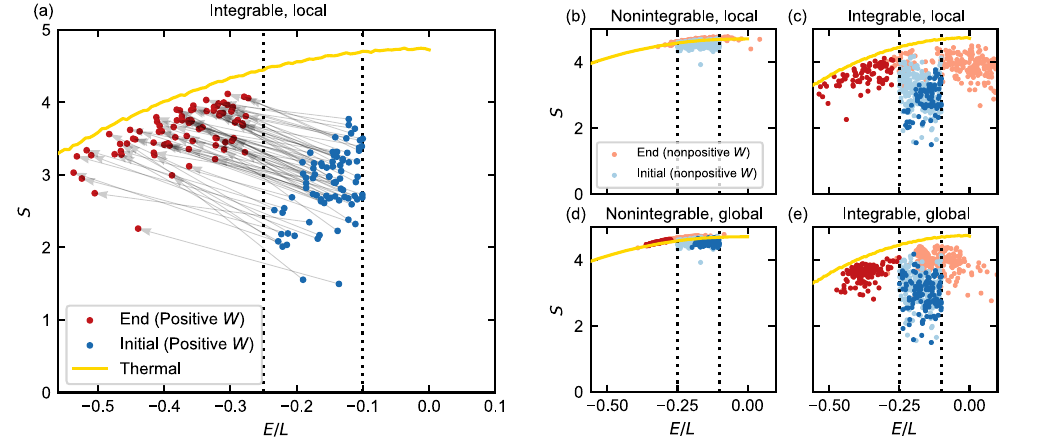}
        \caption{
            The relationship between the work extractability and the EE. 
            The vertical axis represents the half-chain EE $S(\rho_\alpha(t))$, where
            the horizontal axis shows the energy density.
            In panel (a), we show the results for eigenstates with positive work extraction in the integrable case with local control ($k=4$). 
            We observe that the values of EE are all increased.
            Other panels (b)-(e) show results for both positive and negative work extraction. 
            Here each panel represents
            (b)~nonintegrable case under local control~(\(k=4\)), (c)~integrable case under local control~(\(k=4\)), (d)~nonintegrable case under global control~(\(k=L/2\)), and (e)~integrable case under global control~(\(k=L/2\)), respectively.
            All data in this figure correspond to the data at \(L=16\) and \(t=1\) in \fref{fig:Dout_locality}.
            }
        \label{fig:SvsE}
    \end{figure*}
    \begin{figure}[tbp]
        \centering
        \includegraphics[width=\linewidth]{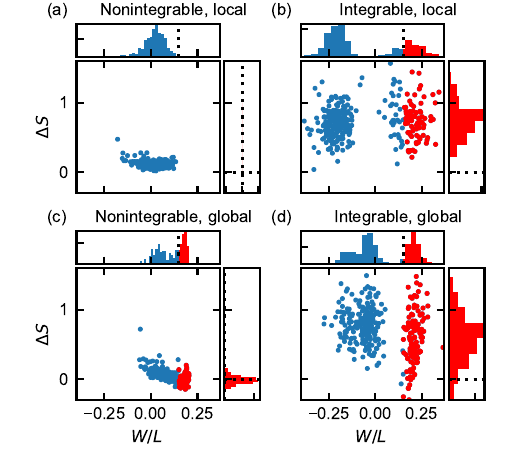}
        \caption{
            The relationship between the EE and the extracted work of the energy eigenstates by the optimized protocols.
            The vertical and horizontal axes represent the changes of the EE and the extracted work density, respectively.
            Each panel shows the result for
            (a)~nonintegrable and local, (b)~integrable and local, (c)~nonintegrable and global, and (d)~integrable and global control protocol. 
            The red markers represent the eigenstates counted as \(D_{\mathrm{pos}}\).
            The histograms on the left of the panels only show the distribution of the eigenstates corresponding to \(D_{\mathrm{pos}}\).
            All data in this figure correspond to the data at \(L=16\) and \(t=1\) in \fref{fig:Dout_locality}.
        }\label{fig:delSvsW}
    \end{figure}
    
    In this study, we adopt as the target Hamiltonian the one-dimensional quantum Ising model under periodic boundary condition:
    \begin{align}
        \hat{H}=\sum_{l=1}^{L}\s_{l}^{z}\s_{l+1}^{z}+h\s_{l}^{z}+g\s_{l}^{x},\label{eq:ising}
    \end{align}
    where $\hat{\sigma}_l^{x,z}$ is the Pauli operator acting on the $l$-th site, $h$ and $g$ are the strength of the longitudinal and transverse magnetic fields, respectively.
    In the subsequent analysis, we exclusively consider \((h,g)=(0.9045,0.809)\) as the nonintegrable case and \((h,g)=(0, 0.5)\) as the integrable case.
    Furthermore, we focus on the energy shell specified from energy density \(E/L\in[-0.25,-0.1]\), and we limit our discussion to
    the zero-momentum sector and the inversion-symmetric sector.

    \subsection{System-size~scaling~of the optimized~work~extraction}\label{sss:size_dep}
        Firstly, we show in \fref{fig:Dout_t} the time evolution of $D_{\rm pos}(t)$ under various control protocols.
        It is consistent with Ref.~\cite{Kaneko2019-ip} that the simple quench dynamics in nonintegrable systems does not extract work at all.
        Meanwhile, as we can see from \fref{fig:Dout_t}(a), the optimization of the control protocol allows us to extract work even in nonintegrable systems.
        It is noteworthy that $D_{\rm pos}(t)$ increases as we increase the locality $k$ of the control protocol. 
        This is somewhat expected behavior, since with larger $k$ we have higher expressibility in the time evolution unitary. 
        As long as the optimization method is successfully performed, we expect $D_{\rm pos}$ to grow with $k$m which is indeed observed in our numerical results. 
        In particular, with $k=8$ we can extract work from the entire energy shell in $L=12$.

        While the gradient-based method and the RL method are different from each other in a sense that the latter only performs discrete optimization, we find that the complex procedure of the RL algorithm allows one to avoid the local minima, so that the optimized protocol for $k=2$ achieves higher $D_{\rm pos}$ by the RL compared to the gradient-based method (see Fig.~\ref{fig:Dout_t}(b)). 
        This is a remarkable benefit of utilizing the RL algorithm, considering the fact that the expressibility of the control unitary is limited to discrete operation set. 
        Note that it is also in agreement with previous works that the RL method is expected to provide a powerful way to determine the quantum control for many-body systems~\cite{Yao2021-jb,Baba2023-iu}. 
        However, here we focus on small or medium-size systems to elaborate on the qualitative difference in the work extractability, and therefore rather focus on the gradient-based method, since it is numerically less demanding for such system sizes.

        Next, we further investigate how the work extractability is affected by the integrability of the Hamiltonian and the locality of control Hamiltonian.
        As is shown in Fig.~\ref{fig:Dout_locality}(a) and (b), the integrability strongly impacts the size scaling when the control operations are local.
        In the nonintegrable case,  we find that \(D_{\mathrm{pos}}\) becomes zero for $L=16$, and further expect that this holds for larger system size $L$ as we further elaborate in the next section.
        In contrast, under global control, \(D_{\mathrm{pos}}\) increases along with the system size $L$ regardless of the integrability, as shown in Fig.~\ref{fig:Dout_locality}(c), (d).
        We summarize such behaviors in Fig.~\ref{fig:Dout_locality}(e) by plotting the size scaling of \(D_{\mathrm{pos}}\) at $t=1$, which increases exponentially except for the nonintegrable system under local operation.

        We remark that the qualitative difference originating from the integrability is exhibited not only under local control operations, but also under global operations as well.
        Namely, we observe that the control time $t$ to achieve some fixed value of $D_{\rm pos}$ remains constant in integrable systems, while it takes longer in nonintegrable systems.
        It is an interesting open question to seek how the required time $t$ scales with the system size.

    \subsection{Work~extractability and athermal~entanglement~entropy}
        To understand the striking difference in the work extractability, we focus on the number of athermal states,
        namely the energy eigenstates whose EE is significantly lower than those of thermal pure quantum states. 
        In nonintegrable systems, it is known that almost all eigenstates are thermal, i.e., the EE converges to that of the canonical ensemble in the thermodynamic limit~\cite{Deutsch2013-in,Beugeling2015-om}.
        Since the thermal EE increases monotonically with energy if one focuses on the energy corresponding to a positive temperature,
        one must reduce the EE to extract work from the system. 
        However, it has been pointed out that, in general it requires exponentially long time under local operation to decrease the EE of the state~\cite{Metz2022-bs,Poulin2011-sb}, and thus there is no efficient way to extract work from any energy eigenstate.
        In contrast, such a property is not generally present in integrable models; there can be exponentially many eigenstates whose EE is lower than the value of thermal EE~\cite{Alba2009-ue,Beugeling2015-om}.
        This means that there is no principle that prohibits one from extracting work from the system even with local operations.

        To examine our conjecture, we analyze the change in the EE before and after the control operation as $\Delta S = S(\rho_\alpha(0)) - S(\rho_\alpha(t))$ where $S$ is the half-chain EE\@.
        In \fref{fig:SvsE}(a), we illustrate that all work-extractable energy eigenstates in integrable system encounter increase in the EE  under local control (\fref{fig:SvsE}(c) also shows data for non-work-extractable states).
        In other words, we are allowed to increase the EE with the control unitary. This does not necessarily require exponentially long control time, and thus expected to be achievable.

        We emphasize that such a scenario is not allowed in nonintegrable systems.
        Figure~\ref{fig:SvsE}(b) shows that, in nonintegrable systems, the EEs of most eigenstates within the energy shell are distributed near the thermal EEs, and the fluctuation is suppressed to be exponentially small~\footnote{In the analysis, we calculate the thermal EE of pure states following \ccite{Lu2019-vt}. The authors of \ccite{Lu2019-vt} calculated the EE of the random pure state with an energy constraint, which they called \emph{ergodic bipartition} state.}.
        This means that, if one desires to extract work, one must reduce the EE of the state by either employing global control, as shown in \fref{fig:SvsE}(d) and (e), or exponentially long unitaries with local terms.

        Figure~\ref{fig:delSvsW} shows the distribution of work and entropy change before and after the control operation of $t=1$. 
        We can again see from Fig.~\ref{fig:delSvsW}(a) and (b) that the EE increases under local operation, and therefore work can be extracted from many eigenstates only in integrable systems.
        In contrast, Fig.~\ref{fig:delSvsW}(c) and (d) reveal the EE can decrease under global operations, and therefore work extraction can be realized even in nonintegrable systems.

        We summarize that these findings are consistent with the proposed scenario that, the qualitative difference in the scaling of athermal eigenstates is directly related with the work extractability in integrable and nonintegrable systems under local operations. 
        Furthermore, these results also imply that we  expect \(D_{\mathrm{pos}}\) to be zero even for \(L>16\) in nonintegrable systems under local control, because the number of athermal eigenstates converges to zero. 
        Meanwhile, this is not the case when one introduces demanding operations such as global operations or exponentially long circuits; we can reduce the EE so that it is possible to extract work.

\section{Conclusion}\label{sse:conclusion}
    In this work, we have utilized numerically optimized quantum control protocols to analyze the work extractability from energy eigenstates of isolated quantum many-body system. Under local control, we find that the integrability is crucial to allow work extraction from exponentially many eigenstates. Conversely, under global control, such a qualitative difference is not observed when the evolution time is sufficiently long. 
    By performing further analysis on the EE, we further find a convincing argument that the work extractability is related with the  number of athermal eigenstates. Namely,  large fluctuation of the EE from the thermal EE in integrable systems is crucial to allow work extraction, while such a mechanism is not present in nonintegrable systems.

    We envision two intriguing future directions. 
    First, it is interesting to explore what is the control time required for positive work extraction. 
    As pointed out in \sref{sss:size_dep}, our fixed duration analysis indicate that positive work extraction requires longer time under larger size in nonintegrable systems. More detailed understanding on the scaling of the control time is an important open problem.
    For instance, one may perform numerical investigation for large-scale systems using variational methods such as tensor network~\cite{vidal2003efficient,vidal2004efficient, daley2004time} or artificial neural networks~\cite{carleo2017solving, yoshioka2019constructing, hartmnann2019neural, nagy2019variational, vicentini2019variational}.
    Another interesting future direction is to study how general our findings hold among various quantum many-body systems.
    We envision that the discrepancy between the integral and nonintegral systems persists under more general local Hamiltonians, which can be naturally expected especially for translationally invariant models.
    While the current study has focused on an one-dimensional model, it is intriguing to explore higher-dimensional systems.


\paragraph*{Acknowledgments.---}
    The authors wish to thank fruitful discussion with Toshihiro Yada.
    S. B. is supported by Materials education program for the future leaders in research, industry, and Technology (MERIT) of The University of Tokyo.
    N.Y. wishes to
    thank JST PRESTO No. JPMJPR2119 and JST Grant Number JPMJPF2221.
    T.S. is supported by JSPS KAKENHI Grant Number JP19H05796,
    JST CREST Grant Number JPMJCR20C1, Japan,
    and JST ERATO-FS Grant Number JPMJER2204, Japan.
    N.Y. and T.S. are also supported by Institute of AI and Beyond of
    The University of Tokyo.
    This work is supported by IBM Quantum.
    The RL is performed on AI Bridging Cloud Infrastructure (ABCI) of National Institute of Advanced Industrial Science and Technology (AIST).

\appendix
\renewcommand\thefigure{\thesection\arabic{figure}}    
\setcounter{figure}{0}
\section{The threshold dependence of the system-size scaling}\label{app:threshold_dep}
    \begin{figure}[tbp]
        \centering
        \includegraphics[width=\linewidth]{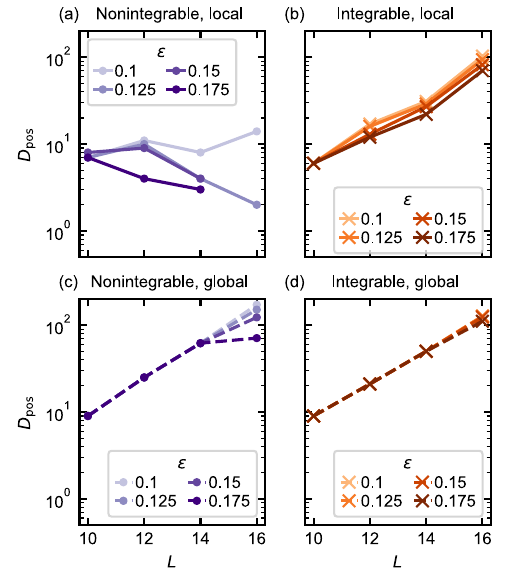}
        \caption{
            The dependence of the threshold \(\varepsilon\) in \eref{eq:dout} on the finite-size scaling of \(D_{\mathrm{pos}}\).
            We fix the parameter in the reward function~\eqref{eq:reward} as the ones used in main part.
            Note that the absence of values corresponding to \(L=16\) in the panel~(a) indicates that \(D_{\mathrm{pos}}\) is zero.
        }\label{fig:dout_dep}
    \end{figure}
    We discuss the relationship between the threshold \(\varepsilon\) in \eref{eq:dout} and the finite-size scaling of \(D_{\mathrm{pos}}\) that is defined in the main text as
    \begin{eqnarray}
        D_{\mathrm{pos}}:=\left|\{\ket{E_{\alpha}}\in\mathrm{shell}|W_{\alpha}(t)\geq \varepsilon L\}\right|.
    \end{eqnarray}
    Note that we fix the parameters in the reward function $a, c, \delta$ \eqref{eq:reward} as the ones used in the main text.

    In Fig.~\ref{fig:dout_dep}, we confirm that the main result as summarized in Table~\ref{teb:outliers_scaling} in the main text is robust under the variation of $\varepsilon.$ 
    While we observe that  $D_{\rm pos}$ seems to saturate for $\varepsilon=0.175$, we argue that this is an artifact due to the vanishing gradient in the reward function in the gradient-based method, rather than the physical phenomena itself. Under the current choice of the reward function, the gradient-based method is valid when $\varepsilon$ does not exceed that of the energy shell width, and therefore we have taken $\varepsilon=0.15$ in the main text.
    
    


\section{Optimization algorithms}\label{app:alg}

\begin{figure}[tb]
    \centering
    \includegraphics[width=\linewidth]{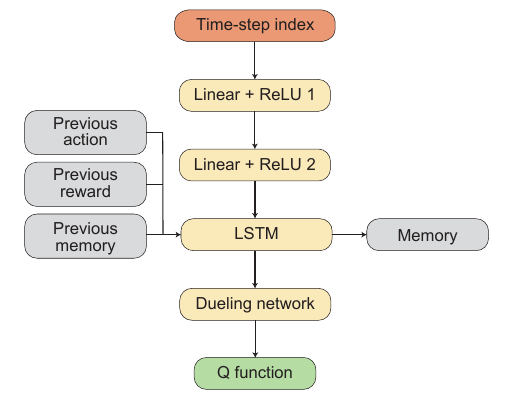}
    \caption{
        The abstract deep NN architecture used for the RL in the present work.
        At each time step \(t_{i}\), the deep NN takes the time-step index \(i\) as input.
        After intermediate computations by fully-connected layers, LSTM~\cite{hochreiter_long_1997}, and the dueling network~\cite{wang_dueling_2016}, the deep NN outputs the estimate of the Q function, from which we determine the next action.
        }\label{fig:RL_architecture}
\end{figure}

\begin{table}[tb]
    \centering
    \begin{tabular}{lr}\toprule
        Linear + ReLU 1 & \(600\rightarrow1024\)\\
        Linear + ReLU 2 & \(1024\rightarrow1024\)\\
        LSTM & \((1024,7,1)\rightarrow 1024\)\\
        Dueling network & \(1024\rightarrow 7\)\\
        \bottomrule
    \end{tabular}
    \caption{
        The Input and Output sizes of each layer are shown in the former and latter of the arrow, respectively. 
        The input of the NN consists of the one-hot representation of the time-step index.
        The input dimension \(600\) corresponds to the number of time step.
        The output dimension \(7\) corresponds to the dimension of the action space.
        }\label{tab:NN_layer}
\end{table} 

\begin{table}[tb]
    \centering
    \begin{tabular}{ll|r}\toprule
        Reward discount \(\eta\)&&0.997\\
        Minibatch size && 608\\
        Sequence length&&40\\
        Optimizer&& Adam~\cite{kingma_adam_2015}\\
        Optimizer setting &learning rate&\(10^{-4}\)\\ 
        &\(\varepsilon\)&\(10^{-3}\)\\
        &\(\beta\)&\((0.9, 0.999)\)\\
        Replay ratio&&1\\
        Gradient norms clip&&80\\
        \bottomrule
    \end{tabular}
    \caption{
        The hyperparameters used in the NNs.
        The agent performs updates on batches of \((\text{minibatch size}\times\text{sequence length})\) observations.
        Replay ratio means the effective number of times each experienced observation is being replayed for the training.
        See \ccite{kapturowski_recurrent_2019} and its previous non-LSTM version, \ccite{horgan_distributed_2018}, for the details of the hyperparameters.
        The other parameters follow the ones in \ccite{stooke_rlpyt_2019}.
        }\label{tab:NN_hyper}
\end{table} 

    \subsection{Gredient-based algorithm}\label{app:greedy}
        We explain the details of the gradient-based algorithm employed in our study.
        We will first discuss how to calculate the coefficients $\{\gamma_i(t)\}$ in the time-dependent Hamiltonian at each time step.
        Next, we introduce the set of operators $\mathcal{B}=\{\hat{O}_i\}$ 
        that constitutes the time-dependent Hamiltonian.
        \subsubsection{Optimizing time-dependent Hamiltonian}

        Given the reward function defined as in Eq.~\eqref{eq:reward} in the main text, we determine the coefficients $\{\gamma_i(t)\}_i$ at each time step $t$ such that the following optimization problem is solved:
            \begin{align}
                \begin{aligned}
                    &\text{maximize}&\frac{dr}{dt}\\
                    &\text{subject to}&\|\ham(t)\|_{2}\leq C.
                \end{aligned}\label{eq:problem}
            \end{align}
            Recall that in the gradient-vased algorithm, the time-evolving Hamiltonian \(\ham(t)\)~\eqref{eq:greedy_ham} in the main text consists of a linear combination of Hermitian operators:
            \begin{align}
                \ham(t)=\sum_{\hO_{i}\in \calB}\gamma_{i}(t)\hO_{i},\label{eq:app_ham}
            \end{align}
            where \(\calB = \{\hat{O}_i\}_i \) denotes a set of Hermitian operators.
            We optimize the coefficients \(\gamma_{i}(t)\) to maximize \(dr/dt\), the derivative of the reward function~\eqref{eq:reward}, and obtain the optimized Hamiltonian at each time step.

            In the following, we derive the explicit representation of the Karush–Kuhn–Tucker (KKT) conditions, the generalization of the Lagrange conditions to the problem subjected to inequality constraints, for the problem defined in Eqs.~\eqref{eq:problem}.
            First, consider the derivative of the reward function~\eqref{eq:reward}:
            \begin{align}
                \frac{dr}{dt}&=\sum_{\alpha}\frac{\partial r}{\partial w_{\alpha}}\frac{d w_{\alpha}}{dt}.\label{eq:deriv_r}
            \end{align}
            The derivative of \(w_{\alpha}\) in \eref{eq:deriv_r} is calculated as
            \begin{align}
                \begin{aligned}[b]
                    \frac{d w_{\alpha}}{dt}=&L^{-1}\frac{d}{dt}\left[ E_{\alpha}-\tr \left[ \ham (\tau)\rho_\alpha(t) \right] \right]\\
                    =&-L^{-1}\tr \left[ \ham (\tau)\frac{dU(t)}{dt}\ket{E_{\alpha}}\bra{E_{\alpha}}U^{\dagger}(t) \right]\\
                    &-L^{-1}\tr \left[ \ham (\tau)U(t)\ket{E_{\alpha}}\bra{E_{\alpha}}\frac{dU^{\dagger}(t)}{dt} \right]\\
                    =&iL^{-1}\tr \left[ \ham (\tau)[\ham (t),\rho_\alpha(t)] \right]\\
                    =&iL^{-1}\sum_{i}\gamma_{i}(t)\tr \left[ \ham (\tau)[\hO_{i},\rho_\alpha(t)] \right]\\
                    =&iL^{-1}\sum_{i}\gamma_{i}(t)\tr \left[ [\ham (\tau),\hO_{i}] \rho_\alpha(t)\right],\label{eq:del_r1}
                \end{aligned}
            \end{align}
            where \([A,B]=AB-BA\) is the commutator, and the substitution of the time-dependent Hamiltonian~\eqref{eq:app_ham} yields the fifth line.
            By substituting \eref{eq:del_r1} into \eref{eq:deriv_r}, the derivative of the reward function reads
            \begin{align}
                \frac{dr}{dt}&=\sum_{i}\gamma_{i}(t)Y_{i}(t),
            \end{align}
            where we have defined
            \begin{align}
                Y_{i}(t):=iL^{-1}\sum_{\alpha}\frac{\partial r}{\partial w_{\alpha}}\tr \left[ [\ham (\tau),\hO_{i}] \rho_\alpha(t)\right].\label{eq:def_Y}
            \end{align}
            Note that the partial derivative of \(r\) in \eref{eq:def_Y} is calculated as
            \begin{align}
                \begin{aligned}[b]
                    \frac{\partial r}{\partial w_{\alpha}}&=\frac{\partial }{\partial w_{\alpha}}\left[ \sigma_{a}(w_{\alpha}-\varepsilon)+c(w_{\alpha}-\delta)\theta(\delta-w_{\alpha}) \right]\\
                    &=\begin{cases}
                        a\sigma_{a}(w_{\alpha}-\varepsilon)\left\{ 1- \sigma_{a}(w_{\alpha}-\varepsilon)\right\}+c &(w_{\alpha}<\delta)\\
                        a\sigma_{a}(w_{\alpha}-\varepsilon)\left\{ 1- \sigma_{a}(w_{\alpha}-\varepsilon)\right\}&(w_{\alpha}\geq\delta),
                    \end{cases}
                \end{aligned}
            \end{align}
            and \eref{eq:def_Y} can be calculated using the expectation values regarding \(\{\rho_\alpha(t)\}_{\alpha}\) at each time step.

            Substituting the time-dependent Hamiltonian~\eqref{eq:app_ham} into the constraint in problem~\eqref{eq:problem} yields
            \begin{align}
                \begin{aligned}[b]
                    C^2&\geq\|\ham(t)\|_{2}^2\\
                    &=\tr \left[ \ham^{\dagger}(t)\ham(t) \right]\\
                    &=\sum_{i,j}\gamma_{i}^{*}(t)\gamma_{j}(t)\tr \left[ \hO_{i}^{\dagger}\hO_{j} \right]\\
                    &=Ld\sum_{i}|\gamma_{i}(t)|^{2},
                \end{aligned}
            \end{align}
            where we assume that $\tr[\hat{O}_i^\dagger \hat{O}_j] = dL\delta_{ij}$.
            As we later see in the next subsection, this assumption is valid for the set $\mathcal{B}$ considered in this work.

            We define the Lagrange function as
            \begin{align}
                \calL:=-\sum_{i}\gamma_{i}(t)Y_{i}(t)+\lambda\left( Ld\sum_{i}|\gamma_{i}(t)|^{2}-C^{2} \right).
            \end{align}
            The KKT condition for the optimization problem~\eqref{eq:problem} is expressed as
            \begin{subequations}
                \begin{align}
                    &\frac{\partial \calL}{\partial \gamma_{i}}=-Y_{i}(t)+2\lambda Ld \gamma_{i}=0\ \text{for any \(i\)},\label{eq:condition_1}\\
                    &Ld\sum_{i}|\gamma_{i}(t)|^{2}-C^{2}\leq 0,\label{eq:condition_2}\\
                    &\lambda\left( Ld\sum_{i}|\gamma_{i}(t)|^{2}-C^{2} \right)=0,\label{eq:condition_3}\\
                    &\lambda\geq 0.\label{eq:condition_4}
                \end{align}\label{eq:condition}
            \end{subequations}
            
            Regarding the constraint, we consider the two cases: (i) \(\|\ham(t)\|_{2}=C\) and (ii) \(\|\ham(t)\|_{2}<C\).
            In the case~(i), the condition~\eqref{eq:condition} becomes
            \begin{subequations}
                \begin{align}
                    & \gamma_{i}=\frac{Y_{i}(t)}{2\lambda Ld}\ \text{for any \(i\)},\label{eq:condition2_1}\\
                    &Ld\sum_{i}|\gamma_{i}(t)|^{2}= C^{2},\label{eq:condition2_2}\\
                    &\lambda\geq 0.\label{eq:condition2_3}
                \end{align}\label{eq:condition2}
            \end{subequations}
            Substituting \eref{eq:condition2_1} into \eref{eq:condition2_2}, we obtain
            \begin{align}
                \lambda=\frac{\|Y(t)\|}{2C\sqrt{Ld}},\label{eq:lambda}
            \end{align}
            where \(\|Y(t)\|=\sqrt{\sum_{i}|Y_{i}(t)|^{2}}\).
            By combining \eref{eq:condition2_1} with \eref{eq:lambda}, the coefficients \(\gamma\) are expressed as
            \begin{align}
                \gamma_{i}(t)=\frac{CY_{i}(t)}{\sqrt{Ld}\|Y(t)\|}.\label{eq:gamma}
            \end{align}
            At each time step, \(\gamma_{i}(t)\) is calculated using the expectation values of \(\ham (\tau)\) and \([\ham (\tau),\hO_{i}]\).

            In the case~(ii), \(\lambda\) must be zero because of \eref{eq:condition_3}.
            By substituting \(\lambda=0\) into \eref{eq:condition_1}, we obtain the condition: \(Y_{i}(t)=0\) for any \(i\).
            This equality holds when \(\rho_\alpha(t)\) commutes with \(\ham (\tau)\) for any \(\alpha\), 
            which results in the vanishing of the gradient of the reward function~(see \eref{eq:del_r1}).
            Note that it is also natural to expect that the case (ii) is essentially not relevant, since the optimization is done such that the gradient is maximized.

            In our analysis, 
            the initial states are the energy eigenstates of \(\ham (0)=\ham (\tau)\).
            Therefore, \(\rho_\alpha(t)\) commutes with \(\ham (\tau)\) for any \(\alpha\), leading to the vanishing of 
            the gradient of the reward function.
            To perform the gradient-based algorithm,
            we add a perturbation to the dynamics at the beginning of the simulation.
            Specifically, we perform a time evolution generated by \(\sum_{l}\s_{l}^{x}\) for a small duration \(\delta t=0.001\).

        \subsubsection{The operator set for gradient-based algorithm}
            In this section, we construct $\mathcal{B}_k$, which is a translationally invariant, spatial inversion-symmetric and orthogonal operator set that act on at most $k$ contiguous sites.

            The first step is to construct a translationally invariant operator set \(\calA_{k}\).
            Consider the $L$-qubit Pauli group $\calP_{L} := \{\pm1, \pm i\} \cdot \{\hat{\sigma}^0, \hat{\sigma}^x, \hat{\sigma}^y, \hat{\sigma}^z\}^{\otimes L}$ where $\hat{\sigma}^{a}~(a=0,x,y,z)$ are Pauli operators.
            For a given Pauli operator $\hat{P}_a \in \mathcal{P}_L$ that acts on at most $k$ contiguous sites, we take a linear combination of translated operators such that \(\hat{Q}_{a}:=\sum_{l=0}^{L-1}\hat{T}^{l}\hat{P}_{a}\hat{T}^{-l}\), where \(\hat{T}\) is one-site translation operator, e.g., \(T\s_{l}^{x}T^{-1}=\s_{l+1}^{x}\).
            Then, from the set of operator \(\{\hat{Q}_{a}\}_{a}\),
            we choose the elements of \(\calA_{k}\) so that there is no duplication. Note that we regard the operators as identical such that only their global phases differ.

            Next, we create an inversion-symmetric operator subset from \(\calA_{k}\), where \(\hat{R}\) is the spatial inversion operator.
            Namely, we remove inversion-asymmetric elements from $\mathcal{A}_k$ and adopt \(\hat{Q}_{a}^{'}=\left( \hat{Q}_{a}+\hat{R}\hat{Q}_{a}\hat{R} \right)/\sqrt{2}\) as the element of $\mathcal{B}_k$  in such a way that there is no redundancy.

            Finally, we confirm that the elements in $\mathcal{B}_k$ satisfy the orthogonality.
            Concretely, we straightforwardly obtain the following:
            \begin{align}
                \begin{aligned}[b]
                    \tr\left[ \hat{Q}_{a}^{\dagger}\hat{Q}_{b} \right]
                    &=\tr\left[ \sum_{l=0}^{L-1}\hat{T}^{l}\hat{P}_{a}^{\dagger}\hat{T}^{-l}\sum_{l'=0}^{L-1}\hat{T}^{l'}\hat{P}_{b}\hat{T}^{-l'} \right]\\
                    &=\sum_{l,l'=0}^{L-1}\tr\left[\hat{P}_{a}^{\dagger}\hat{T}^{-(l-l')}\hat{P}_{b}\hat{T}^{l-l'} \right] \\
                    &=\sum_{l,l'=0}^{L-1}d\delta_{a,b}
                    =\begin{cases}
                        Ld &(a=b)\\
                        0&(a\neq b).
                    \end{cases}
                \end{aligned}                    
            \end{align}
            Here, we used the fact that we have chosen $\hat{Q}_a \in \mathcal{A}_k$ such that there is no redundancy, which implies that $\hat{P}_a$ does not coincide with any $\hat{P}_{b\neq a}$ under any translation operation.
            Following similar calculation, we also confirm that the inversion-symmetrized elements \(\hat{Q}_{a}^{'}\) satisfy the orthogonality and the norm \(\|\hat{Q}_{a}^{'}\|_{2}=\sqrt{Ld}\). 
            As a result, we have verified the orthogonality of $\mathcal{B}_k$ and the norm of elements \(\|\hat{Q}_{a}\|_{2}=\|\hat{Q}_{a}^{'}\|_{2}=\sqrt{Ld}\).

    \subsection{Deep reinforcement learning}\label{app:RL}
        When adopting deep RL, we construct the protocol with a unitary sequence, in which each element corresponds to the time evolution at each time step, chosen from a fixed set of unitaries.
        The elements of the fixed set of unitaries are generated by the following Hermitian operator set \(\{ \ham_{m} \}_{m}\):
        \begin{align}
            \begin{aligned}
                &\sum_{l}J\sigma^z_l\sigma^z_{l+1}+h_{\mathrm{I}}\sigma^z_l,\ 
                \sum_{l}J\sigma^z_l\sigma^z_{l+1}+h_{\mathrm{N}}\sigma^z_l,\\
                &\sum_{l}\s_{l}^{x}\s_{l+1}^{y}+\s_{l}^{y}\s_{l+1}^{x},\ 
                \sum_{l}\s_{l}^{y}\s_{l+1}^{z}+\s_{l}^{z}\s_{l+1}^{y},\\
                &\sum_{l}g_{\mathrm{I}}\sigma^x_l,\ 
                \sum_{l}g_{\mathrm{N}}\sigma^x_l\ 
                \sum_{l}\s_{l}^{y},
            \end{aligned}
        \end{align}
        where \((J,\ h_{\mathrm{I}},\ h_{\mathrm{N}},\ g_{\mathrm{I}},\ g_{\mathrm{N}})=(1,\ 0,\ 0.9045,\ 0.5,\ 0.809)\).
        Note that the norm of these terms satisfies the upper bound \(\sqrt{2Ld}\) adopted in the main text.
        These terms, which are linear combinations of elements in \(\calB_{2}\), are also used in \ccite{Yao2021-jb,Baba2023-iu}.
        In \sref{sss:size_dep}, we set the time-step length as \(0.04\) and the number of time-step is \(600\).

        Deep reinforcement learning, specifically deep Q-learning, utilizes a NN to approximate the following optimal action-value function Q~\cite{mnih_human-level_2015,sutton_reinforcement_2018}:
        \begin{align}
            Q^{*}(t_{i},m)=\max_{\pi}\mathbb{E}_{\pi}\left\lbrack r_{t_{i}}+\sum_{n=1}^{\infty}\eta^{n} r_{t_{i+n}} \middle|\ham(t_{i})=\ham_{m},\ \pi \right\rbrack,\label{eq:Q}
        \end{align}
        which denotes the maximum sum of rewards \(r_{t}\) discounted by \(\eta\) (\(0<\eta<1\)) in a stochastic policy 
        that selects actions based on a probability distribution as \(\pi\left\lparen  m|t_{i} \right\rparen=\Pr\left\lparen  m|t_{i} \right\rparen\).
        A potent variant of Q-learning harnesses the capabilities of deep NNs to represent the action-value function, which is hence referred to as deep RL algorithm~\cite{Li2018-cl,Henderson2017-aw}.
        
        In this paper, we direct our focus toward a non-distributed implementation~\cite{stooke_rlpyt_2019} of a deep RL algorithm termed R2D2~\cite{kapturowski_recurrent_2019}.
        R2D2 is a form of deep Q-learning algorithm, and assumes that the agent can obtain partial information about the state of the environment.
        Figure~\ref{fig:RL_architecture} shows the overall picture.
        The details regarding the network structure are shown in  Table~\ref{tab:NN_layer} and Table~\ref{tab:NN_hyper}. 
%
    
\end{document}